\documentclass[12pt,epsf]{article}
\usepackage{epsf}
\begin{document}
\newcommand{\sheptitle}
{LMA MSW Solution from the Inverted Hierarchical Model of Neutrinos}
\newcommand{\shepauthor}
{Mahadev Patgiri$^{\dag}$ and N. Nimai Singh$^{\ddag}$\footnote{Corresponding
author.\\{\it{E-mail address}}: nimai@guphys.cjb.net}}
\newcommand{\shepaddress}
{$^{\dag}$ Department of Physics, Cotton College, Guwahati-781001, India\\
$^{\ddag}$ Department of Physics, Gauhati University, Guwahati-781014, India}
\newcommand{\shepabstract}
{We examine  whether the inverted hierarchical model of neutrinos is compatible
 with the explanation of the large mixing angle (LMA)MSW solution of the solar neutrino problem. The
left-handed Majorana neutrino mass matrix for the inverted hierarchical model, is 
generated through the seesaw mechanism using the diagonal form of the Dirac neutrino 
mass matrix and the non-diagonal texture of the right-handed Majorana mass matrix. 
In a model independent way we construct a specific form of the charged lepton mass matrix having a special 
structure in 1-2 block, which   contribution to the leptonic mixing (MNS) matrix leads to the 
predictions $sin^{2}2\theta_{12}=0.8517$, $sin^{2}2\theta_{23}=0.9494$ and $|V_{e3}|=0.159$ at 
the unification  scale. These predictions  are found to be  
consistent with the LMA MSW solution of the solar neutrino problem. 
The inverted hierarchical model is also found to be stable
 against the  quantum radiative corrections in the MSSM.  A  numerical analysis of the renormalisation 
group equations (RGEs) in the MSSM  shows a mild decrease of the mixing angles
 with the decrease of energy scale and the corresponding values of the neutrino mixings at the top-quark 
mass scale are $sin^{2}2\theta_{12}=0.8472$, $sin^{2}2\theta_{23}=0.9399$ and 
$|V_{e3}|=0.1509$ respectively.}
\begin{titlepage}
\begin{flushright}
hep-ph/0112123
\end{flushright}
\begin{center}
{\large{\bf\sheptitle}} 
\bigskip\\
\shepauthor
\\
\mbox{}
\\
{\it\shepaddress}
\\
\vspace{.5in}
{\bf Abstract}\bigskip\end{center}\setcounter{page}{0}
\shepabstract
\end{titlepage}
\section{Introduction}
Neutrino physics is one of the fast developing areas of particle physics. The recent  Super-Kamiokande 
experimental results on both solar[1] and atmospheric[2] neutrino oscillations support the approximate 
bimaximal mixings.  Though these results favour the large mixing angle (LMA) MSW solution with active 
neutrinos, such interpretation is not beyond doubt  at this stage[3,4,5]. We also have little idea about 
the pattern of the neutrino mass spectrum whether it is hierachical or inverted hierarchical, 
and  both possibilities are consistent with the neutrino oscillation explanations of the atmospheric 
and solar neutrino deficits[5,6]. The data 
from the long baseline experiment using a Neutrino factory will be able to confirm the actual 
  pattern  of the neutrino masses in the near future[7]. 

In the theoretical front the hierarchical model of neutrino masses and its generation have been widely
 studied and  found to be consistent  with the explanation of the LMA MSW solar neutrino solution[8,9].
 However,  the inverted hierarchical model of neutrino masses   generally predicts the maximal mixing 
angles $\theta^{\nu}_{12}$ and $\theta^{\nu}_{23}$  close to $45^{0}$, and  are  suitable for the 
explanation of the vacuum oscillation (VO) solution of the solar neutrino oscillation[6,10] 
and   the  atmospheric neutrino oscillation data. The atmospheric data  gives   the lower bound  at 
$sin^{2}2\theta_{23}\geq 0.88$  and the best-fit value at  $\Delta m^{2}_{23}=3 \times 10^{-2}eV^{2}$. 
It is quite obvious that the prediction from the inverted hierarchical model  fails to  explain the 
LMA MSW solution which has upper experimental limit[4,6] $sin^{2}2\theta_{12}\leq 0.988$ at $95\%$ C.L., 
and  the best-fit values  $sin^{2}2\theta_{12}=0.8163$ and $\Delta m^{2}_{12}=4.2 \times 10^{-5}eV^{2}$. 
 Combining LMA MSW solution and atmospheric data, the best-fit value of the 
mass splitting parameter is obtained[6] as $\xi=\Delta m^{2}_{12}/\Delta m^{2}_{23}=0.014$.
It has been argued [6,10] that  the diagonalisation of the charged lepton mass matrix cannot give 
a significant contribution  to
 $\theta^{\nu}_{12}$ needed for the explanation of the LMA MSW solution. On such ground the inverted 
hierarchical models are assumed to be inconsistent with the LMA MSW solution. An  attempt  was  made to 
explain the LMA MSW solution from the  inverted hierarchical model by considering two types of 
charged lepton mass matrices[11] and was partially successful.  We are interested here to make 
further investigations
  in this paper   whether the inverted hierarchical model  gives an acceptable LMA MSW solution when we 
include  the contribution from the diagonalisation of the charged lepton mass matrix having special 
form in the 1-2 block, to the leptonic mixing matrix. 

The paper is organised as follows. In section 2 we  outline  the seesaw mechanism  for the generation 
of the neutrino mass matrix which leads to the  inverted hierarchical mass pattern, and 
the contruction of the charged lepton mass matrix suitable for the LMA MSW solution. In section 3 
we describe briefly the procedure for the analysis of the renormalisation group equations (RGEs) 
within the minimal supersymmetric standard model (MSSM). This is followed by a summary and discussion 
in section 4.   

\section{Generation of the inverted hierarchical neutrino mass matrix and the charged lepton mass matrix}
The inverted hierarchical model of neutrinos  has its origin from 
the low energy non-seesaw models[12], e.g., the Zee-type of model using a singly charged singlet 
scalar field and also the models with an approximate conserved
$L_{e}-L_{\mu}-L_{\tau}$ lepton number. However, it is also  possible  to generate the inverted 
hierarchical model  through the seesaw mechanism at high energy scale within the framework of the 
grand unified theories with a chiral U(1) family symmetry[10,11]. In a model independent way, 
we consider the Dirac neutrino mass matrix $m_{LR}$ and the non-diagonal form of the right-handed 
Majorana mass matrix $M_{R}$ in the seesaw formula [13] given by 
\begin {equation}
m_{LL}=m_{LR}M^{-1}_{R}m_{LR}^{T}
\end{equation} 
where $m_{LL}$ is the left-handed Majorana mass matrix. The leptonic mixing matrix known as 
 the MNS mixing matrix [14] is defined  by
\begin{equation}
V_{MNS}= V_{eL}V^{\dag}_{\nu L}
\end{equation}
where $V_{eL}$ and $V_{\nu L}$ are obtained from the diagonalisation of the charged lepton and $m_{LL}$ as
$$m^{diag}_{e}= V_{eL} m_{e} V^{\dag}_{eR}$$
\begin{equation}
m^{diag}_{LL}= V_{\nu L} m_{LL} V^{\dag}_{\nu L}
\end{equation}
If the charged lepton mass matrix is diagonal, the MNS matrix (2) is simply given by
\begin{equation}
V_{MNS}= V^{\dag}_{\nu L}
\end{equation}
We can always express $m_{LL}$ in the basis where the charged lepton mass matrix is  diagonal,
$$m^{\prime}_{LL}= V_{eL} m_{LL} V^{\dag}_{eL},$$
 $$m^{\prime  diag}_{LL}= V^{\prime}_{\nu L} m^{\prime}_{LL} V^{\prime \dag}_{eL},$$
\begin{equation}
V_{MNS}= V^{\prime \dag}_{\nu L}
\end{equation}

From the above expressions we can calculate  the experimentally determined  quntities as follows:\\
 (i) the neutrino mass splitting parameter, $\xi= \frac{|\Delta m^{2}_{12}|}{|\Delta m^{2}_{23}|}$\\
(ii) the atmospheric mixing angle, $S_{at}= \sin^{2}2\theta_{23}= 4 |V_{\mu 3}|^{2}\left(1- |V_{\mu 3}|^{2}\right)$\\
(iii) the solar mixing angle, $S_{sol}= \sin^{2}2\theta_{12}= 4 |V_{e2}|^{2} |V_{e1}|^{2}$\\
(iv) the CHOOZ angle,  $S_{C}= 4 |V_{e3}|^{2}\left(1- |V_{e3}|^{2}\right)$ or simply $|V_{e3}|$.\\
 The $V_{fi}$ where  $f=\tau, \mu, e$ and   $i=1, 2, 3$,  are the elements of the MNS mixing matrix.

First, we consider the diagonal form of the charged lepton mass matrix $m_{e}$ given by  
\begin{equation}
m_{e}=
\left(\begin{array}{ccc}
\lambda^{6}& 0 & 0 \\
0 & \lambda^{2} & 0 \\
0 & 0 & 1
\end{array}\right) m_{\tau}
\end{equation}
where the Wolfenstein parameter[15] $\lambda=0.22$ and the ratios of the charged lepton masses are 
$m_{\tau}:m_{\mu}:m_{e}= 1:\lambda^{2}:\lambda^{6}$ respectively. From Eq.(6) we get  
$V_{eL}=1$ and $V_{MNS}= V^{\dag}_{\nu L}$ as in Eq.(4).
Again we consider the diagonal form of the Dirac neutrino mass matrix $m_{LR}$ 
as the up-quark mass matrix[16],
\begin{equation}
m_{LR}=
\left(\begin{array}{ccc}
\lambda^{8}& 0 & 0 \\
0 & \lambda^{4} & 0 \\
0 & 0 & 1
\end{array}\right) m_{t}
\end{equation}
where the up-quark masses are in the ratios[17]  $m_{t}:m_{c}:m_{u}= 1:\lambda^{4}:\lambda^{8}$.
Now, the proper choice of the elements in $M_{R}$, enables us to generate the inverted hierarchical
neutrino mass matrix. We present here  the following examples[18]:\\ 

\underline{\it Example (a)}
\begin{equation}
m_{LL}=
\left(\begin{array}{ccc}
0 & 1 & 1 \\
1 & \lambda^{3} & 0 \\
1 & 0 & \lambda^{3}
\end{array}\right) m_{0},
\end{equation}
with the choice
$$M_{R}=
\left(\begin{array}{ccc}
-\lambda^{22}& \lambda^{15} & \lambda^{11} \\
\lambda^{15} & \lambda^{8} & -\lambda^{4} \\
\lambda^{11}& -\lambda^{4} & 1
\end{array}\right) v_{R}$$
Eq.(8) yields
\begin{equation}
V_{MNS}=V^{\dag}_{\nu L}=
\left(\begin{array}{ccc}
0.70577 & 0.70844 & -1.11 \times 10^{-16} \\
-0.50094 & 0.49906 & -0.70711 \\
-0.50094 & 0.49906 & 0.70711
\end{array}\right) v_{R}
\end{equation}
and the neutrino mass eigenvalues $m_{i}= (1.4195, 1.4089, 0.0105)m_{0}$, $i= 1, 2, 3.$  This gives the mass splitting parameter $\xi= \Delta m^{2}_{12}/\Delta m^{2}_{23}=0.014$,
and the mixing angles $\sin^{2}2\theta_{12}=0.9999$, $\sin^{2}2\theta_{23}\approx 1.00$,  $|V_{e3}|=0.0$.
\\

\underline{\it Example (b)}
\begin{equation}
m_{LL}=
\left(\begin{array}{ccc}
0 & 1 & 1 \\
1 & -(\lambda^{3}-\lambda^{4})/2 & -(\lambda^{3}+\lambda^{4})/2 \\
1 & -(\lambda^{3}+ \lambda^{4})/2 &-(\lambda^{3}- \lambda^{4})/2
\end{array}\right) m_{0},
\end{equation}
with the choice 
$$M_{R}=
\left(\begin{array}{ccc}
\lambda^{23}& \lambda^{16} & \lambda^{12} \\
\lambda^{16} & \lambda^{8} & -\lambda^{4} \\
\lambda^{12}& -\lambda^{4} & 1
\end{array}\right) v_{R}$$
leading to $m_{i}= (1.4195, 1.4089, 0.00239)m_{0}$, $\xi=0.0151$, $\sin^{2}2\theta_{12}=0.9999$, 
$\sin^{2}2\theta_{23}\approx 1.00$, $|V_{e3}|=0.0$.
\\

\underline{\it Example (c)}
\begin{equation}
m_{LL}=
\left(\begin{array}{ccc}
\lambda^{3} & 1 & 1 \\
1 & \lambda^{4}/2 & -\lambda^{4}/2 \\
1 & -\lambda^{4}/2 & \lambda^{4}/2
\end{array}\right) m_{0},
\end{equation}
with the choice
$$M_{R}=
\left(\begin{array}{ccc}
0 & \lambda^{16} & \lambda^{12} \\
\lambda^{16} & \lambda^{8} & -(\lambda^{4}+\lambda^{12}) \\
\lambda^{12}& -(\lambda^{4}+\lambda^{12}) & 1
\end{array}\right) v_{R}$$
leading to $m_{i}= (1.4195, 1.4089, 0.002343)m_{0}$, $\xi=0.015$, $\sin^{2}2\theta_{12}=0.9999$, 
$\sin^{2}2\theta_{23}\approx 1.00$, $|V_{e3}|=0.0$.

In the  above results the $V_{MNS}$ obtained from the  $m_{LL}$ alone fails to  explain the 
LMA MSW solution, and any small deviation in $m_{LL}$ will hardly affect $\sin^{2}2\theta_{12}$[6,10]. 
The last hope is that there could be a significant
contribution to $\theta_{12}$ from the diagonalisation of the charged lepton mass matrix having 
special structure in 1-2 block [11].  We wish to  examine how
$\theta_{sol}= ( \theta^{\nu}_{12}- \theta^{e}_{12})$ can resolve the LMA MSW solar neutrino 
mixing scenario [11].

We parametrise the charged leptonic mixing $V_{eL}$ by the following three rotations[19,20]
$$V_{eL}= \bar{R}_{23}\bar{R}_{13}\bar{R}_{12}$$
\begin{equation}      
=\left(\begin{array}{ccc}
1 & 0 &0\\ 0 & \bar{c}_{23} & \bar{s}_{23}\\ 0 & -\bar{s}_{23}& \bar{c}_{23}
\end{array}\right)
\left(\begin{array}{ccc}
\bar{c}_{13} & 0 & \bar{s}_{13}\\ 0 & 1 & 0 \\ -\bar{s}_{13} & 0 & \bar{c}_{13}
\end{array}\right)
\left(\begin{array}{ccc}
\bar{c}_{12} & \bar{s}_{12} & 0 \\ -\bar{s}_{12} & \bar{c}_{12} & 0 \\ 0 & 0 & 1
\end{array}\right)
\end{equation}
where $\bar{s}_{ij}=\sin\theta^{e}_{ij}$ and $\bar{c}_{ij}=\cos\theta^{e}_{ij}$. Putting $\theta^{e}_{13}=
\theta^{e}_{23}=0$, Eq.(12) reduces to 
\begin{equation}
V_{eL}=
\left(\begin{array}{ccc}
\bar{c}_{12} & \bar{s}_{12} & 0 \\ -\bar{s}_{12} & \bar{c}_{12} & 0 \\ 0 & 0 & 1
\end{array}\right)
\end{equation}
This gives a special form in the 1-2 block. We then reconstruct[19] the charged lepton mass matrix
using Eq.(13) from  the relation,
$$m_{e}= V^{\dag}_{eL}m^{diag}_{e}V_{eR}$$
\begin{equation} 
=\left(\begin{array}{ccc}
\lambda^{6}\bar{c}_{12}^{2}+\lambda^{2}\bar{s}_{12}^{2} &
 \lambda^{6}\bar{c}_{12}\bar{s}_{12}-\lambda^{2}\bar{c}_{12}\bar{s}_{12} & 0 \\ 
\lambda^{6}\bar{c}_{12}\bar{s}_{12}-\lambda^{2}\bar{c}_{12}\bar{s}_{12} &
\lambda^{6}\bar{c}_{12}^{2}+\lambda^{2}\bar{s}_{12}^{2}  & 0 \\ 
0 & 0 & 1
\end{array}\right)
\end{equation}
For a specific choice of $\theta^{e}_{12}= 13^{0}$, and $\lambda= 0.22$,  Eq.(14) leads to 
\begin{equation}
m_{e}=
\left(\begin{array}{ccc}
0.00256 & -0.01058 & 0\\
-0.01058 & 0.04596 & 0 \\
0 & 0 & 1
\end{array}\right)
\end{equation}
which has a special form in the 1-2 block.
The diagonalisation of $m_{e}$ in Eq.(15) gives 
\begin{equation}
V_{eL}=
\left(\begin{array}{ccc}
-0.97439 & -0.22488 & 0\\
-0.22488 & 0.97439 & 0 \\
0 & 0 & 1
\end{array}\right)
\end{equation}
 which is now completely unitary. 
The corresponding eigenvalues of the charged lepton mass matrix are given by 
\begin{equation}
m_{e}^{diag}=( 1.182\times 10^{-4}, 4.84\times 10^{-2}, 1.0)m_{\tau}
\end{equation}
which gives almost correct physical mass ratios[17] $m_{e}:m_{\mu}:m_{\tau}= \lambda^{6}:\lambda^{2}:1$.
The MNS mixing matrix (2) is now calculated, using Eqs.(9) and (16), as
\begin{equation}
V_{MNS}=V_{eL}V_{\nu L}^{\dag}=
\left(\begin{array}{ccc}
-0.5750 & -0.8025 &0.1590 \\
-0.6468 & 0.32696 & 0.6890 \\
-0.50094 & 0.49906 & 0.70711
\end{array}\right)
\end{equation}
This  leads to the mixing angles $sin^{2}2\theta_{12}=0.8517$, $sin^{2}2\theta_{23}=0.9494$, and
 $|V_{e3}|=0.159$, and these predictions  are consistent with the explanation of LMA MSW solution. 
The possible  choice 
of $\theta^{e}_{12}=14^{0}$ in Eq.(14) also leads to the predictions of $sin^{2}2\theta_{12}=0.8298$, 
$sin^{2}2\theta_{23}=0.9415$, and $|V_{e3}|=0.1710$ while maintaining the good prediction of the  
ratios of the charged lepton masses. However its $|V_{e3}|$ is above the CHOOZ and PALO VERDE 
experimental constraint [21] of  
$|V_{e3}|\leq 0.16$. 

Taking the first result with  $\theta^{e}_{12}=13^{0}$, the left-handed neutrino mass  $m^{\prime}_{LL}$  
in the basis where the charged lepton mass matrix is diagonal(5), is now expressed for our convenience, as
\begin{equation}
m^{\prime}_{LL}
=\left(\begin{array}{ccc}
0.437972 & -0.897698 &-0.973193 \\
-0.897698 & -0.443296 & -0.230068 \\
-0.973193 & -0.230068 & -0.005324
\end{array}\right)m_{0}
\end{equation}
where $V_{MNS}= V^{\prime \dag}_{\nu L}$ is same as earlier given in Eq.(18),
 and the neutrino mass eigenvalues are 
$$m_{i}=\left(1.4196, 1.4089, 4.234 \times 10^{-7}\right)m_{0};    i=1, 2, 3$$
which give  the mass splitting parameter,  $\xi=\Delta m^{2}_{12}/\Delta m^{2}_{23}= 0.014.$

\section{Renormalisation effects in MSSM}
It is desirable to inspect how the values of $sin^{2}2\theta_{12}$, $sin^{2}2\theta_{23}$, $|V_{e3}|$  
and $\xi$ evaluated at the unification scale where the seesaw mechanism is operative, 
respond to the renormalisation
group analysis on running from higher scale $\left(M_{u}=2\times 10^{16}GeV\right)$ down to the top quark 
mass scale $(\mu=m_{t})$ [22]. We consider the renormalisation group equations (RGEs) for the three gauge
couplings $(g_{1}, g_{2}, g_{3})$ and the third family Yukawa couplings $(h_{t}, h_{b}, h_{\tau})$
in the minimal supersymmetric standard model (MSSM) in the standard fashion[23]. At high scale 
$\mu=2\times 10^{16}GeV$, we assume the unification of gauge couplings as well as third generation
Yukawa couplings for large $tan\beta$ [23]. We choose the input $\alpha_{2}= (5/3)\alpha_{1}=\alpha_{3}=1/24,$ and $h_{t}=h_{b}=h_{\tau}=0.7$ corresponding to large $tan\beta=v_{u}/v_{d}$.

We express $m_{LL}$ in terms of K, the coefficient of the dimension 
five neutrino mass operator[24,25,26,27] in a scale-dependent manner,
\begin{equation}
m_{LL}(t)= v^{2}_{u}(t) K(t)
\end {equation}
where $t=ln(\mu)$ and $v_{u}(t)$ is the scale-dependent[27] vacuum expectation value (VEV) 
$v_{u}= v_{0} sin\beta$, $v_{0}= 174 GeV$. In the basis where the charged lepton mass matrix is diagonal,
we can write Eq.(20) as [25,27]
\begin{equation}
m^{\prime}_{LL}(t)= v^{2}_{u}(t) K^{\prime}(t)
\end {equation}
where $K^{\prime}(t)$ is the coefficient of the dimension five neutrino mass operators in the basis 
where the charged lepton mass matrix is diagonal.  The evolution equations are given by[27] 
\begin{equation}
\frac{d}{dt}lnv_{u}(t)=
\frac{1}{16\pi^{2}}\left[\frac{3}{20} g^{2}_{1}+ \frac{3}{4} g^{2}_{2} -3 h^{2}_{t}\right],
\end{equation}
\begin{equation}
\frac{d}{dt}lnK^{\prime}(t)=
-\frac{1}{16\pi^{2}}\left[\frac{6}{5} g^{2}_{1}+ 6 g^{2}_{2} - 6 h^{2}_{t}-h^{2}_{\tau}\delta_{i3}-h^{2}_{\tau}\delta_{3j}\right],
\end{equation}
The evolution equation of $m^{\prime}_{LL}(t)$ in Eq.(21) simplifies[27] to 
\begin{equation}
\frac{d}{dt}lnm^{\prime}_{LL}(t)=
\frac{1}{16\pi^{2}}\left[-\frac{9}{10} g^{2}_{1}- \frac{9}{2} g^{2}_{2} +h^{2}_{\tau}\delta_{i3}+h^{2}_{\tau}\delta_{3j}\right].
\end{equation}
Upon integration from high scale $(t_{u}=\ln M_{u})$ to lower scale
 $(t_{0}=\ln m_{t})$ where $t_{0}\le t\le t_{u}$
and $t=\ln \mu$, we get[27]
$$m^{\prime}_{LL}(t_{0})= e^{\frac{9}{10}I_{g_{1}}(t_{0})} e^{\frac{9}{2}I_{g_{2}}(t_{0})}$$
\begin{equation}
\times \left(\begin{array}{ccc}
m^{\prime}_{LL11}(t_{u}) & m^{\prime}_{LL12}(t_{u})  & m^{\prime}_{LL13}(t_{u}) e^{-I_{\tau}(t_{0})}\\
m^{\prime}_{LL21}(t_{u}) & m^{\prime}_{LL22}(t_{u})  & m^{\prime}_{LL23}(t_{u}) e^{-I_{\tau}(t_{0})} \\
 m^{\prime}_{LL31}(t_{u}) e^{-I_{\tau}(t_{0})} & m^{\prime}_{LL32}(t_{u}) e^{-I_{\tau}(t_{0})}  
& m^{\prime}_{LL23}(t_{u}) e^{-2I_{\tau}(t_{0})}
\end{array}\right)
\end{equation}
where
$$I_{g_{i}}(t_{0})=\frac{1}{16\pi^{2}}\int^{t_{u}}_{t_{0}} g^{2}_{i}(t) dt, 
i=1, 2, 3;$$ 
\begin{equation}
I_{f}(t_{0})=\frac{1}{16\pi^{2}}\int^{t_{u}}_{t_{0}} h^{2}_{f}(t) dt ,  f= t, b, \tau.
\end{equation}
Using the numerical values of $I_{g_{i}}(t)$ and $I_{f}(t)$ at different energy scales 
$t$, $t_{0}\le t\le t_{u}$
the left-handed Majorana mass matrix $m^{\prime}_{LL}(t)$ in Eq.(25) is estimated at different energy 
scales from the value of $m^{\prime}_{LL}(t_{u})$ given in Eq.(19).
At each scale the leptonic mixing matrix $V_{MNS}(t)=V^{\prime \dag}_{\nu L}(t)$ is calculated and 
this in turn, gives mixing angles $sin^{2}2\theta_{12}$,   $sin^{2}2\theta_{23}$ and $|V_{e3}|$. 
For example, at the top-quark mass scale  $t_{0}=\ln m_{t}=5.349$, we have calculated 
$I_{\tau}(t_{0})=0.100317$ and the leptonic mixing matrix
\begin{equation}
V_{MNS}=
\left(\begin{array}{ccc}
-0.56962 & -0.80795 &0.15085 \\
-0.67061 & 0.35075 &- 0.65364 \\
-0.4752 & 0.47349 & 0.74161
\end{array}\right)
\end{equation}
which leads to the low-energy predictions:  $sin^{2}2\theta_{12}=0.8472$ and  $sin^{2}2\theta_{23}=0.9399$. There is a mild  reduction
from the values estimated at the high energy scale $t_{u}$. This feature shows the compatibility of the inverted hierarchical model
with the explanation of the LMA MSW solution. The parameter $|V_{e3}|=0.15085$ meets the CHOOZ constraint  
$|V_{e3}|\le 0.16$ [21]. The neutrino mass eigevalues at low-energy scale are 
 obtained as $m_{i}=(1.3533, 1.3436, 3.8376\times 10^{-7})m_{0}$.  However the mass splitting parameter 
 $\xi=\Delta m^{2}_{12}/\Delta m^{2}_{23}=0.01449$ remains almost constant.
The running of the mixing angles $S_{sol}=sin^{2}2\theta_{12}$ and $S_{at}=sin^{2}2\theta_{23}$ is  
shown in Fig.1 by the solid-line and dotted-line respectively. Both parameters decrease with decrease in 
energy scale $t$. This is a desirable feature for maintaining the stability condition of the inverted 
hierarchical model. Note that there is no exponential term which depends  on the top-quark Yukawa coupling 
integration in Eq.(25). This differs from the expressions  calculated in earlier papers[11,25]. The absence
 of such term increases the stability criteria of the mass matrix at low $tan\beta$ region and this is made 
possible only when we consider the scale-dependent vacuum expectation value in Eq.(22). 
As discussed before,  if the CHOOZ constraint is  relaxed to $|V_{e3}|\le.2$, then we would be able to 
get even lower  value of $sin^{2}2\theta_{12}$ suitable for the explanation of the best-fit value of 
 the LMA MSW solution. 
\\
\vbox{
\noindent
\hfil
\vbox{
\epsfxsize=12cm
\epsffile[130 380 510 735]{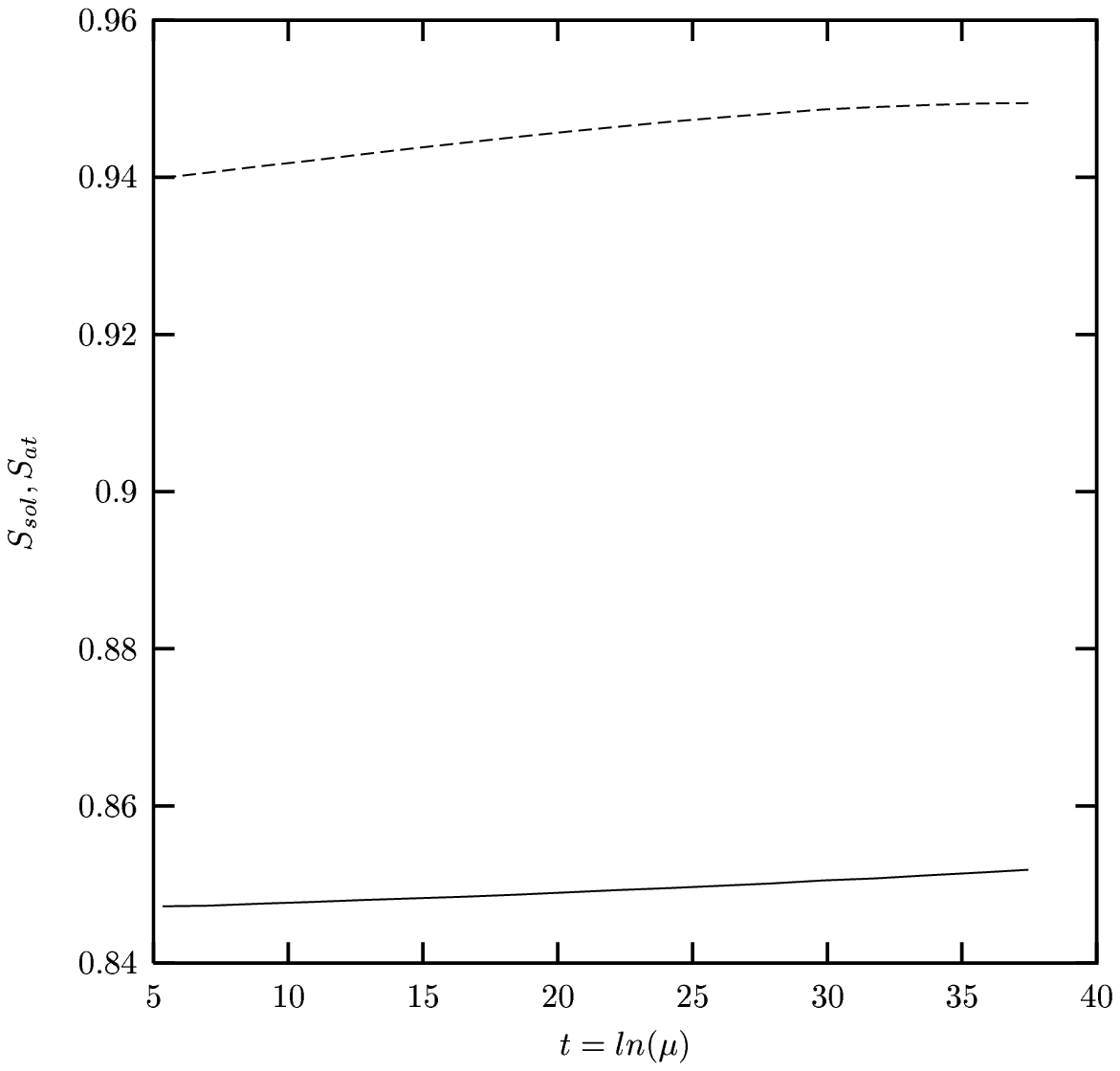}}

{\narrower\narrower\footnotesize\noindent
{Fig.1.}
 Variations  of $S_{sol}=sin^{2}2\theta_{12}$ and $S_{at}=sin^{2}2\theta_{23}$ with energy scale 
$t=\ln(\mu)$, which 
are represented by solid-line and dotted-line, respectively.
\par\bigskip}}
\\

We now briefly discuss the analytic solution for the evolution of the neutrino mixings in MSSM. The equation for the evolution of the neutrino mixing angle $\sin^{2}2\theta_{ij}$  can be approximated
as two flavour mixing  in terms of the neutrino mass eigenvalues[26] $m_{1}, m_{2},  m_{3}$ by a generalised evolution equation,
\begin{equation}
16\pi^{2}\frac{d}{dt}\sin^{2}2\theta_{ij}=-2\sin^{2}2\theta_{ij}(1-\sin^{2}2\theta_{ij})
( h^{2}_{j}-h^{2}_{i})\frac{m_{j}+m_{i}}{m_{j}-m_{i}}
\end{equation}
where $i<j$ and $i, j= 1,2,3;$ or  $e,\mu,\tau$. 
For solar mixing angle $\theta_{12}$, we have 
\begin{equation}
16\pi^{2}\frac{d}{dt}\sin^{2}2\theta_{12}=-2\sin^{2}2\theta_{12}(1-\sin^{2}2\theta_{12})
( h^{2}_{\mu}-h^{2}_{e})\frac{m_{2}+m_{1}}{m_{2}-m_{1}}
\end{equation}
and for atmospheric mixing angle $\theta_{23}$ we have
\begin{equation}
16\pi^{2}\frac{d}{dt}\sin^{2}2\theta_{23}=-2\sin^{2}2\theta_{23}(1-\sin^{2}2\theta_{23})
( h^{2}_{\tau}-h^{2}_{\mu})\frac{m_{3}+m_{2}}{m_{3}-m_{2}}
\end{equation}
Since $h^{2}_{\tau}>h^{2}_{\mu}>h^{2}_{e}$ and $m_{1}>m_{2}>m_{3}$, both   $S_{sol}$ and $S_{at}$ in Eqs.(29) and (30)
increase with increasing energy scale. These features are plainly 
visible in the Fig.1.

A few more comments on our choice of the texture of the charged lepton mass matrix are also presented.
Here we  examine the forms of the texture of the charged lepton mass matrix $m_{e}$ and its diagonalisation 
matrix $V_{eL}$ in Eqs.(15)and (16). For simplicity of our analysis
 we express them in the following approximate forms 
\begin{equation}
m_{e}\sim
\left(\begin{array}{ccc}
\lambda^{4} & -\lambda^{3} & 0 \\
-\lambda^{3} & \lambda^{2} & 0 \\
0            &    0        & 1 
\end{array}\right)
\end{equation}
and\\
\begin{equation}
V_{eL}\sim
\left(\begin{array}{ccc}
-1 & -\lambda & 0 \\
-\lambda & 1 & 0 \\
0            &    0        & 1 
\end{array}\right)
\end{equation}
It is interesting to note that the position of the  zeros in the mass matrix in Eq.(31) have the same
 structure with those of lepton mass matrix obtained by Ibanez and Ross[29] in the gauge theory of the 
standard model with an horizontal $U(1)$ gauge factor.
Such form of the texture of the charged lepton mass matrix is also 
proposed by  Georgi and Jarlskog[30] in $SU(5)$ model, and this  can be realised in a model based on $SUSY SO(10)\times U(2)$ using a 
$126$-dimensional Higgs[31].
The CKM matrix of the quark mixings  defined by $V_{CKM}=V_{uL}V_{dL}^{\dag}$, is usually  parametrised 
by[15] 
\begin{equation}
V_{CKM}\sim
\left(\begin{array}{ccc}
1-\lambda^{2}/2  & \lambda & A \lambda^{3}(\rho -i\eta) \\
-\lambda & 1-\lambda^{2}/2 & A\lambda^{2} \\
A \lambda^{3}(1-\rho-i\eta)& -A \lambda^{2} & 1 
\end{array}\right)
\end{equation}
where $\lambda=0.22$ and  $|A|=0.90$. For our choice of the diagonal up-quark mass matrix in Eq.(7),
 we have $V_{uL}=1$ leading to 
 $V_{CKM}=V_{dL}^{\dag}$. 
 Since  $m_{d}=m_{e}^{T}$, we have $V_{eL}=V_{dL}^{\dag}=V_{CKM}$. Neglecting higher power of $\lambda$ 
in Eq.(33), we have
\begin{equation}
V_{eL}\sim
\left(\begin{array}{ccc}
1  & \lambda & 0 \\
-\lambda & 1 & 0 \\
0& 0 & 1 
\end{array}\right)
\end{equation}
which is almost same as  $V_{eL}$ given in Eq.(32) except the difference in sign before some entries. The positions of the zeros are the same. Such linkage gives partial justification to  our motivation for  the choice of the charged lepton mass matrix (15).  

\section{Summary and Discussion}
The left-handed Majorana neutrino mass matrix $m_{LL}$ which  explains the inverted hierarchical
pattern of neutrino masses, has been generated from the seesaw mechanism using non-diagonal texture
of the right-handed Majorana neutrino mass matrix $M_{R}$ and diagonal form of the Dirac neutrino 
mass matrix . We have explained the leptonic mixing matrix generated from the diagonalisation 
of $m_{LL}$ of 
the inverted hierarachical model and the mixing angles so far obtained  $sin^{2}2\theta_{12}\approx 0.999$, is too large for the explanation of the LMA MSW solution. Such high value of  $sin^{2}2\theta_{12}$ can 
be tonned down by considering the contribution from the charged lepton mass matrix having special structure 
in the 1-2 block. With such consideration, the predictions  on the mixing angles at the high scale are $sin^{2}2\theta_{12}=0.8517$, $sin^{2}2\theta_{23}=0.9494$ and $|V_{e3}|=0.159$ which are consistent with the LMA MSW solution.

The above results which are calculated at the high energy scale (say, $\mu=M_{u}=2\times10^{16}GeV$) where the seesaw mechanism operates, decrease with the decrease in energy scale, under the quantum radiative  corrections within the framework of the MSSM. This is a good feature at least for $\sin^{2}2\theta_{12}$ in this inverted hierarchical model as it gives the stability under quantum radiative corrections and shows complete consistency of the model with 
the explanation of the LMA MSW solution. The present finding fails to  support the claim that an arbitrary mixing at the high scale can get ``magnified'' to a large mixing, and even possibly maximal mixing at the low scale [28]. 
Experimental data from a Neutrino  factory may confirm the pattern of the neutrino masses in near future, and hence the sign 
of $\Delta m^{2}_{23}$.
 Such confirmation of the detailed pattern of neutrino masses and their  mixing angles is  very important 
as it may give a clue to the understanding of quark masses and their mixing angles within the 
framework of an all-encompassing theory[5].
 
Though we have constructed both $m_{LL}$ and $m_{e}$ in a model independent way and have shown how the inverted hierarchical model of neutrinos can explain the present experimental data particularly LMA MSW solution, 
the present work is expected to be an  important clue for building  models from the grand unified 
theories with the chiral U(1) symmetry.

\end{document}